\documentclass[a4paper]{article}
\usepackage[UKenglish]{babel}
\usepackage{a4wide}

\usepackage{amsmath,amssymb,amsthm}
\usepackage{gensymb}
\usepackage{graphicx}

\usepackage{textcomp}
\usepackage{paralist}
\usepackage[lined,boxed,commentsnumbered]{algorithm2e}
\usepackage{tikz}

\newcommand{\abs}[1]{\left\vert#1\right\vert}
\let\div\relax\DeclareMathOperator{\div}{div}
\newcommand{\cI}{\mathcal{I}}

\newcommand{\cS}{\mathcal{S}}

\newcommand{\expten}[1]{\cdot 10^{#1}}
\newcommand{\ind}[1]{\chi_{#1}}
\newcommand{\R}{\mathbb{R}}
\newcommand{\unit}[1]{\mathrm{#1}}
\newcommand{\vect}[1]{\textbf{#1}}

\theoremstyle{remark}\newtheorem*{remark}{Remark}

\graphicspath{{../}{figures/}} 

\title{From individual behaviour to an evaluation of the collective evolution of crowds along footbridges}
\author{	Luca Bruno \\
		{\small\it Politecnico di Torino} \\
		{\small\it Department of Architecture and Design} \\
		{\small\it Viale Mattioli 39, 10125, Torino, Italy} \\[5mm]
		Alessandro Corbetta\thanks{The work of this author is supported by a Lagrange Foundation Ph.D. scholarship.} \\
		{\small\it Politecnico di Torino} \\
		{\small\it Department of Structural and Geotechnical Engineering} \\
		{\small\it Corso Duca degli Abruzzi 24, 10129 Torino, Italy}\\
                {\small\it and}\\
                {\small\it Eindhoven University of Technology} \\
		{\small\it CASA\,-\,Centre for Analysis, Scientific computing and Applications}\\
                {\small\it Department of Mathematics and Computer Science} \\
		{\small\it P.O. Box 513, 5600 MB Eindhoven, The Netherlands} \\[5mm]
		Andrea Tosin \\
		{\small\it Consiglio Nazionale delle Ricerche} \\
		{\small\it Istituto per le Applicazioni del Calcolo ``M. Picone''} \\		
		{\small\it Via dei Taurini 19, 00185 Rome, Italy}
		}
\date{}

\begin{document}

\maketitle

\begin{abstract}
This paper proposes a crowd dynamic macroscopic model grounded on microscopic phenomenological observations which are upscaled by means of a formal mathematical procedure.

The actual applicability of the model to real world problems is tested by considering  the pedestrian traffic along footbridges, of interest for Structural and Transportation Engineering. The genuinely macroscopic quantitative description of the crowd flow directly matches the engineering need of bulk results. However, three issues beyond the sole modelling are of primary importance: the pedestrian inflow conditions, the numerical approximation of the equations for non trivial footbridge geometries, and the calibration of the free parameters of the model on the basis of \textit{in situ} measurements currently available. These issues are discussed  and a solution strategy is proposed.

\medskip


\end{abstract}

\section{Introduction}
\label{Sec:footbr-intro}
The study of the dynamics of crowds defines a wide research field that, in recent times, is knowing an increasing expansion fostered by several scientific communities such as Applied Mathematics~\cite{cristiani2014BOOK}, Physics~\cite{helbing2001RMP}, Biomechanics~\cite{zajaca_2003}, Cognitive Psychology~\cite{warren_2006}, Computer and Graphics Visualisation~\cite{xu_2014}, Safety Engineering~\cite{Gwynne_1999,Zheng_2010}, Transportation Engineering~\cite{Duives_2013} and Structural Engineering~\cite{ingo_2012,ziv_jsv_2005}. Correspondingly, a number of engineering applications and traffic conditions have been addressed, e.g.~evacuation under panic conditions, passenger flow in rush hour conditions or pedestrian loads in crowd events. The reader can refer to the reported review papers for a complete surveys of the state of the art in each context.

Despite the fact that these different scientific fields are trying to model the same physical entity, i.e.~flows of crowds of pedestrians, research ideas have evolved almost independently. Among the 1734 bibliographic references discussed in the selected surveys~\cite{Duives_2013,helbing2001RMP,ingo_2012,warren_2006,xu_2014,zajaca_2003,Zheng_2010} only 1 author is jointly cited in 6 of them, 2 authors in 5 scientific fields, 10 authors in 3 surveys and 69 in at least 2 surveys. In other terms, less than 5\% of the references are somewhat shared among the different disciplines. Hence, it is reasonable to believe that each discipline has developed perspectives and techniques that are characteristic of their own.

Furthermore, genuine multidisciplinary surveys are scarce up to now and do not include all fields reported above (examples are~\cite{ali_2013,venuti2009PLR}). In this  study, which develops in the context of the  traffic of pedestrians along footbridges, we aim at combining priorities of two fields by joining a formal mathematical approach with practical requirements of Structural Engineering. In particular, we deal with the three following working issues, that in general can be  related to the need of complementing rigorous  mathematical procedures with  practical requirements of engineering problems:

\begin{enumerate}
\item \textit{Modelling scale.} Crowd models are often categorized via their representation scale, commonly microscopic or macroscopic (cf.~e.g.~\cite{helbing2001RMP} and~\cite[Chapter 4]{cristiani2014BOOK}). This partition is indeed not exhaustive as kinetic~\cite{agnelli2015M3AS,degond2013KRM,degond2013JSP} or fully discrete, cellular automata-based, models are in use as well~\cite{blue_1998}. 

Microscopic models, adopting the point of view of single pedestrians, are the closest to the scale of the single individuals at which phenomena generate. The pioneering ``social force'' model~\cite{helb_1995} provides an example. Beyond fundamental purposes, this approach has been used in applied perspective (e.g.~analysis of evacuation~\cite{schadschneider2011EEE,zheng2009modeling} or of the traffic along footbridges~\cite{Carroll20122685}) and  is the foundation of different  engineering  commercial software packages (e.g. Steps\copyright ~ by Mott MacDonald or MassMotion\copyright ~ by Arup, see also~\cite{haron_2012}). Although widely adopted, this approach features a crucial con: it depends on (social) force terms involving a large number of free parameters difficult to calibrate~(see, e.g.~\cite{corbetta2015parameter,johansson2007specification,zanlungo2011social}). Conversely, from the seminal paper~\cite{hughes2000flc}, macroscopic models are scarcely supported by phenomenological assumptions, and their parameters are usually estimated from statistical observations like the so-called fundamental diagram~\cite{daa}. These models are usually adopted in fundamental studies to depict the emerging behaviour of crowd traffic, e.g. stop-and-go waves and clogging at exits~\cite{Twarogo_2014}. To the Authors' best knowledge, no commercial code is based on  macroscopic framework models, nonetheless engineering applications can be found (e.g. in the case of footbridges~\cite{bruno2009JSV,ven_2013}). Mesoscopic (kinetic) models, adopting a statistical point of view on the microscopic states of the pedestrians, see e.g.,~\cite{agnelli2015M3AS,degond2013KRM}, demand even greater computational efforts and costs. In fact, for two-dimensional simulations they require to handle four independent variables (two components for both the spatial position and the velocity) plus time.

\item \textit{Bulk descriptors.}
Engineering applications are usually interested in bulk, possibly statistical, descriptors of crowd scenarios, rather than in the microscopic ``configuration'' of each pedestrian. Examples of such descriptors are the Level Of Service (LOS)~\cite{fruin1987BOOK} in Transportation Engineering, the evacuation time in Safety Engineering, and the crowd load in Structural Engineering. Microscopic models allow one to evaluate these descriptors as by-product of many-particles simulations, usually having high computational costs. However, a macroscopic model  provides ``natively'' averaged quantities such as the density, and hence the LOS, avoiding unnecessary computations. Nevertheless, up to now, this potential pro has not been exploited. 

\item \textit{Actual applicability.}
According to the Authors, the lack of popularity of the macroscopic approach in Engineering  is probably related to the difficulties in handling three technical issues, not at the core of a purely modelling perspective:
\begin{itemize}
\item the  definition of proper inflow and inner boundary conditions, to replicate, respectively, time-dependent incoming flows and the effects of different types of environmental walls. To our best knowledge,  fundamental studies usually focused on unbounded geometries~\cite{cristiani2011MMS,piccoli2009CMT}, while more applied studies have adopted standard inflow conditions~\cite{algadhi1990modelling,algadhi1991simulation,colombo2011modelling};
\item a numerical approximation of the governing partial differential equations in two-dimen\-sional domains having articulated geometries. To our best knowledge, only few studies have been devoted to this issue, e.g.~\cite{Huang_2009,Twarogo_2014,xia_2008}.
\item the calibration of the model parameters on the basis of available  \textit{in situ} measurements;
\end{itemize}
\end{enumerate}

In view of these issues, in this study we derive a macroscopic crowd model that yields ready-to-use macroscopic information in direction of engineering needs (issue~2). The model is based on behavioural considerations at the scale of the single pedestrian, which we ``upscale'' via  a formal mathematical procedure (issue~1). Furthermore, we give a proof-of-concept of its applicability (issue~3), addressing the case of pedestrian flows across footbridges; in particular
\begin{itemize}
\item the model parameters are conceived to be calibrated on the basis of few bulk experimental data, considered in the current monitoring practice; 
\item the inflow boundary condition aims at reflecting common working conditions for footbridges. The footbridge is initially empty and pedestrians enter  according to a queuing process, depending on both external conditions and pedestrian dynamics at the footbridge entrance;
\item the repulsive effects of different types of walkway parapets (inner boundary) are taken into account by the model;
\item the footbridge non trivial geometry by means of a numerical scheme employing unstructured triangular grids.
\end{itemize}

The paper is organized in four more sections that reflect the conceptual approach  described above. In particular, in Section~\ref{sect:crowdmod} we propose the phenomenological model and we deduce the mathematical model therefrom; in Section~\ref{sect:real_events} we discuss further elements required to simulate real world crowd events along footbridges; in Section~\ref{sect:application} we consider results from simulations of crowd events in four computational domains inspired by real world geometries as well as the calibration of the model. The paper is closed with the discussion in Section~\ref{sect:conclusions}.

\section{Crowd modelling}
\label{sect:crowdmod}
In this section we introduce and discuss the crowd model. We first deduce a phenomenological model describing the dynamics of a representative ``test'' pedestrian in a moving crowd (Section~\ref{sect:phenmod}), then we obtain a mathematical model dealing with the evolution of the overall collectivity (Section~\ref{sect:mathmod}).

In the modelling approach that we propose, individual pedestrians determine their velocity on the basis of interactions they have with the neighboring distribution of pedestrians. For the sake of simplicity, we treat the pedestrian-collectivity interaction mechanism via a velocity term called \emph{interaction velocity}. The interaction velocity plays the role of a perturbation of a second velocity component, the \emph{desired velocity}, which, instead, models the velocity one would keep in the absence of other nearby agents. The sum of desired velocity and interaction velocity gives rise to the pedestrian \emph{total velocity}. The final model is not new \textit{per se} and has been introduced and discussed in~\cite{cristiani2011MMS,piccoli2009CMT,piccoli2011ARMA}. Nonetheless, it has always be postulated \textit{as is}, apart from a deduction based on the dynamics of the individuals which we here discuss. Moreover, here, for the first time, we consider the model in the direction of its practical application in view of the issues mentioned in Section~\ref{Sec:footbr-intro}.

\subsection{Phenomenological modelling of pedestrian dynamics}
\label{sect:phenmod}
Let $D\subseteq\R^d$ be the spatial domain where the crowd walks. Usually $d=2$, but the presented approach is sufficiently general to allow one to keep the dimension $d$ generic. In the current section, $D$  coincides with the whole unbounded space $\R^d$. Nonetheless, in Section~\ref{sect:real_events} we will relax this requirement allowing bounded domains, in the spirit of  engineering needs.

At time $t>0$, we treat the individual point of view in terms of the spatial position\footnote{Throughout the paper, the subscript $t$ is used to denote a dependence on time (hence, in particular, not a time derivative).} $X_t\in D$ of a generic representative walker, that we refer  to as the \emph{test pedestrian}. On the other hand, we express the collective point of view in terms of a \emph{crowd distribution} represented via the pedestrian density function (integrable at all times, unit: $\unit{ped/m^2}$) $\rho_t:D\to[0,\,+\infty)$, that satisfies 
\begin{equation}\label{eq:crowding}
	\int_{E}\rho_t(x)\,dx=\textup{measure of the \emph{crowding} of $E$ at time $t$},
\end{equation}
for any choice of a (Borel) measurable $E\subset D$ see Fig.~\ref{fig:ind_vs_coll}.

We assume pedestrians to have a desired walk velocity, which is kept in the absence of other people nearby. We express this velocity  through a vector field 
\begin{equation*}
	v_d:D\rightarrow\R^d,
\end{equation*}
that is evaluated at the agent's position. On the other hand, to obtain the \emph{interaction velocity} we consider a vector-valued interaction kernel
\begin{equation*}
	K:\R^d\rightarrow\R^d,
\end{equation*}
such that $K(y-X_t)$ models the reaction that the test pedestrian in $X_t$ has to another pedestrian in $y$ on the basis of their relative position $y-X_t$. However, since pedestrians generally interact with   few nearby walkers from the surrounding crowd distribution, the interaction velocity is ultimately constructed by considering the action of $K$ only in a region $S(X_t)\subset D$ around $X_t$ and weighting pairwise interactions  by the local crowding $\rho_t$ (cf. analogous considerations in~\cite{lachapelle2011mean}), i.e.,
\begin{equation}
	v_i[\rho_t](X_t)=\int_{S(X_t)}K(y-X_t)\rho_t(y)\,dy,
	\label{eq:vi}
\end{equation}
The set $S(X_t)$ is the so-called \emph{sensory region} of the test pedestrian (see Fig.~\ref{fig:ind_vs_coll} and, e.g.,~\cite{bruno2011AMM,fruin1987BOOK}). This region is expected to be bounded and possibly anisotropic with respect to the pedestrian direction of movement. 

\newcommand{\fontsizehereP}{\large}
\begin{figure}[t]
\centering
\begin{tikzpicture}
\node[inner sep=0pt] at (-2.3,2){
\includegraphics[width=0.5\textwidth]{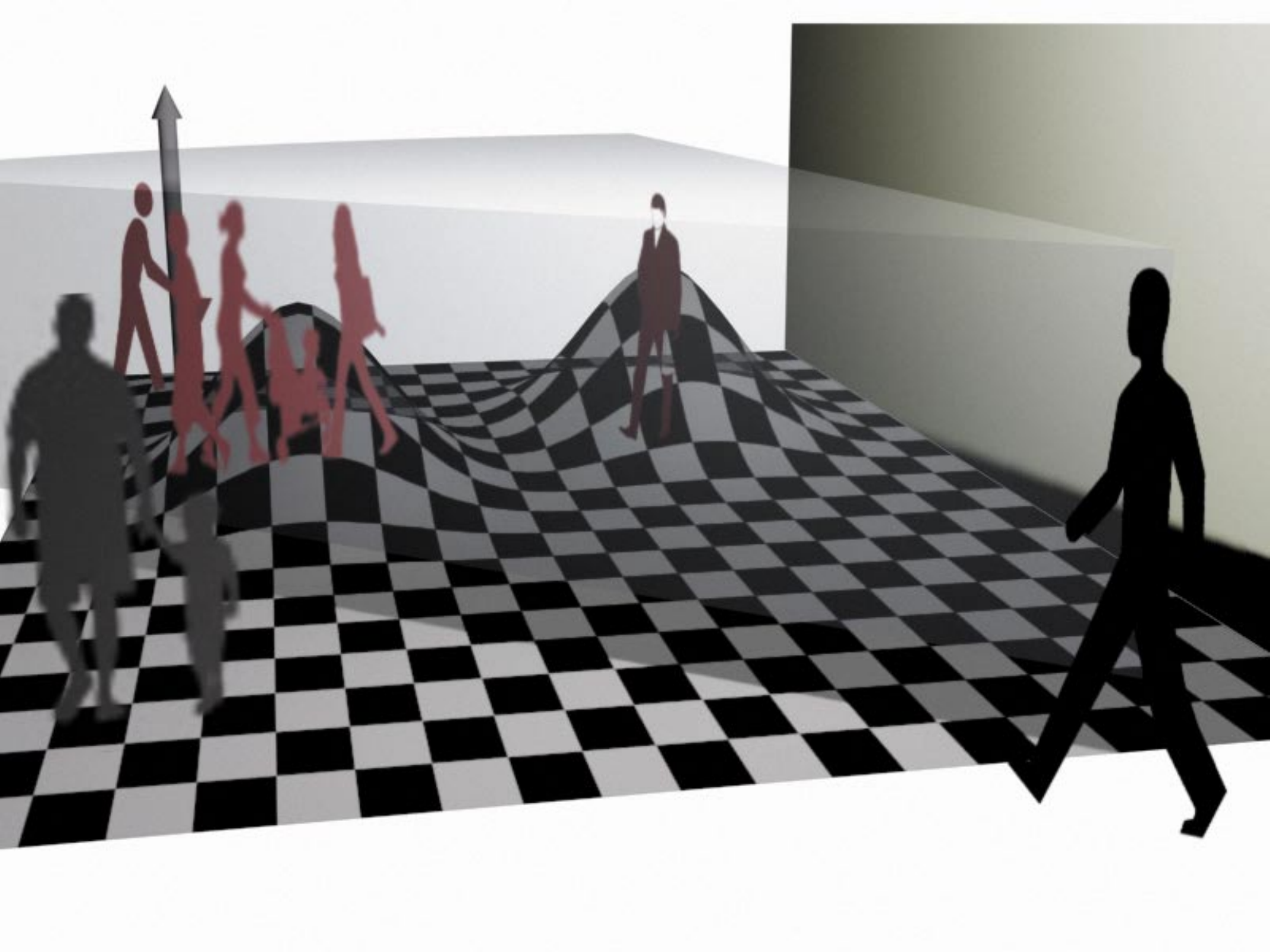}};
\node at (-.65,3.5) {\fontsizehereP $S(X_t)$}; 
\node at (.17,.09) {\fontsizehereP $X_t$};
\draw[fill=black] (0,.43) circle (2pt);
\node at (-5.6,.85) {\fontsizehereP $\rho_t$}; 
\end{tikzpicture}
\caption{The test pedestrian facing the crowd distribution ahead (checker texture)}
\label{fig:ind_vs_coll}
\end{figure}
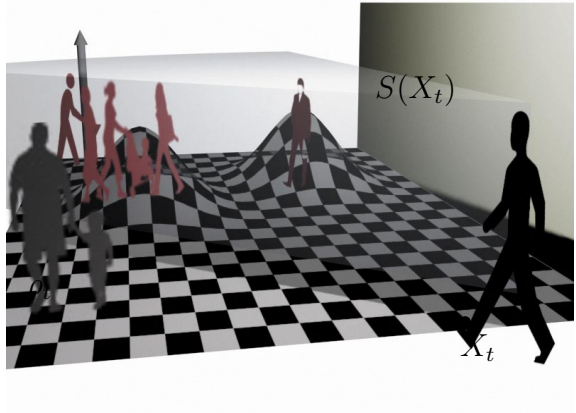

To obtain the total velocity, we add the interaction velocity  to the desired velocity, that gives
\begin{equation}
	 \dot{X}_t=v_d(X_t)+v_i[\rho_t](X_t)=v_d(X_t)+\int_{S(X_t)}K(y-X_t)\rho_t(y)\,dy,
	\label{eq:streamlines}
\end{equation}
which models the trajectory the test pedestrian follows locally  and details how $v_i$ induces deviations from the direction indicated by $v_d$.

The model we deduced thus far features two state variables, $X_t$ and $\rho_t$, nonetheless we provided just one evolution equation~\eqref{eq:streamlines}. We obtain the second equation as follows. In order to be consistent with its definition of measure of crowding, $\rho_t$ has to be a \emph{material quantity} for pedestrians, which means that the initial crowd distribution $\rho_0$ is transported in space and time by the pedestrians' motion. This  is expressed by the relation
\begin{equation}
	\int_{X_t(E)}\rho_t(x)\,dx=\int_{E}\rho_0(\xi)\,d\xi,\qquad \forall t>0, \quad\forall\,E\subseteq D.
	\label{eq:pushfwd}
\end{equation}
Note that $X_t$ is here viewed as a \emph{flow map} $X_t:D\to D$ such that $x=X_t(\xi)$ is the position occupied at time $t$ by the (test) walker that initially was in $\xi$. 

\subsection{Mathematical model}
\label{sect:mathmod}
In this section we derive a self-consistent mathematical model from the coupled evolution equations~\eqref{eq:streamlines}-\eqref{eq:pushfwd}. Proceeding formally like in the Reynolds Transport Theorem (cf.,~e.g.,~\cite{Marsdenbook}), we expand the time derivative of the material quantity  $\rho_t$. Given a region $X_t(E)$ which follows the pedestrian flow, by the Change of Variables Theorem, we can write
\begin{align*}
	\frac{d}{dt}\int_{X_t(E)}\rho_t(x)\,dx &= \frac{d}{dt}\int_{E}\rho_t(X_t(\xi))J_t(\xi)\,d\xi, \\
	\intertext{where we have made the substitution $x=X_t(\xi)$ (and $J_t(\xi)$ is the Jacobian of the flow map $X_t$, i.e.,
		$J_t(\xi)=\abs{\det{\frac{\partial X_t(\xi)}{\partial\xi}}}$, ``$\det$'' standing for the determinant of a matrix).
		Computing now the time derivative under the integral sign we get:}
	&= \int_{E}\Bigl(\partial_t\rho_t(X_t(\xi))J_t(\xi)+\nabla\rho_t(X_t(\xi))\cdot\dot{X}_t(\xi)J_t(\xi) \\
	&\phantom{=}\qquad +\rho_t(X_t(\xi))\dot{J}_t(\xi)\Bigr)\,d\xi
	\intertext{whence, recalling that $\dot{J}_t(\xi)=J_t(\xi)\div{\dot{X}_t(\xi)}$,}
	&= \int_{E}\left(\partial_t\rho_t(X_t(\xi))+\div{[\rho_t(X_t(\xi))\dot{X}_t(\xi)]}\right)J_t(\xi)\,d\xi,
\end{align*}
where ``$\div{}$'' is the divergence operator. Since the right-hand side of~\eqref{eq:pushfwd} is time-independent, the expression above must be zero. Using~\eqref{eq:streamlines} and substituting back $x=X_t(\xi)$ we finally obtain
\begin{equation}\label{eq:integral}
	0=\int_{X_t(E)}\left\{\partial_t\rho_t(x)+\div{\left[\rho_t(x)\bigl(v_d(x)+v_i[\rho_t](x)\bigr)\right]}\right\}\,dx.
\end{equation}
Since~\eqref{eq:integral} is true for every choice of the sub-domain $E$, under the assumption of regularity of $X_t$ (when understood as a flow) and of $\rho_t$,   the following conservation law  holds
\begin{equation}
	\partial_t\rho_t+\div{\left(\rho_t\left(v_d+v_i[\rho_t]\right)\right)}=0.
	\label{eq:strong}
\end{equation}
 One can have well-posedness of the Cauchy problem obtained by complementing~\eqref{eq:strong} with an integrable initial condition 
\begin{equation}
 \rho_0\geq 0 \quad \textup{at\ } t=0 \textup{\ in\ } D.
	\label{eq:initcond}
\end{equation}
The interested reader is referred to~\cite{piccoli2013AAM,tosin2011NHM} for technical details.

Equation~\eqref{eq:strong} expresses the conservation of the ``number of pedestrians''. Indeed it is formally a conservation law for the quantity $\rho_t$ featuring a nonlocal flux due to the interaction velocity $v_i[\rho_t]$. Other macroscopic crowd models are constructed by postulating a closure of the continuity equation~\eqref{eq:strong} by means of  fundamental diagrams, i.e.  relationships between the density and the velocity of pedestrians (see, e.g.,~\cite{colombo2012M3AS}). In our case, we do not rely on fundamental diagrams, but rather we model directly the interaction dynamics among pedestrians. We refer the interested reader to~\cite{bruno2011AMM} for further models of pedestrian dynamics based on interactions, possibly leading to nonlocal fluxes.

\begin{remark}
It is worth pointing out that the desired velocity field $v_d(x)$ is assumed to be the same for all pedestrians, in the sense that any individual passing through the point $x\in D$ at any time has the same instantaneous desired velocity. In other words, $v_d$ is given as an Eulerian field, in such a way that everyone walks locally in the same direction and heads toward the same destination. In general, one may expect that the domain $D$ is possibly populated by different groups of people with locally different desired directions of movement. For instance, in the case of a footbridge, there may be people walking, say, leftward and others walking, say, rightward. In order to model such a scenario, it is necessary to introduce the concept of \emph{subpopulations} within the crowd. In practice, one describes the crowd by means of two densities, $\rho^1$ and $\rho^2$, each of which satisfies an equation analogous to~\eqref{eq:strong} but with two different desired velocity fields $v_d^1$, $v_d^2$ pointing in the proper direction. Clearly, either population is insensitive to the desired velocity of the other population, but the two populations interact with one another. Therefore the interaction velocity of each population has to include an additional term similar to~\eqref{eq:vi}, which takes into account cross-interactions with the members of the other population. These concepts have been already developed for our model~\eqref{eq:strong} in previous works, see e.g.,~\cite[Chapter 5]{cristiani2014BOOK} and~\cite{colombi2015JMB} for an example not directly related to crowd dynamics but sharing with the present context the same modelling approach. In footbridge applications, no computational simulation exist which accounts practically for bidirectional crowd flows, while only one experimental campaign~\cite{ziv_2012} discerns traffic directions. On the contrary, a few simulations~\cite{bruno2009JSV,Carroll20122685,ven_2013} and \textit{in situ} observations \cite{dal,fujino1993EESD,setareh_2011} report of footbridges which typically exhibit unidirectional crowd flows, for instance in case of opening ceremonies, public demonstrations or festive events, end of sport events. For this reason, we believe that a single-population model with just one desired velocity field is enough for our purposes.
\end{remark}

\section{Model applicability: proof of concept}
\label{sect:real_events}
In this section we obtain a tool to analyse crowd events in real domains on the basis of the modelling framework previously deduced. From now on the dimension of the domain is fixed to $d=2$.

First, to obtain a self-consistent tool we have to assign an expression to $K$ (cf.~\eqref{eq:vi}). To meet engineering needs, this expression is expected to be synthetic, although obtained on a phenomenological basis.

Second, since we focus on crowd flows developing on footbridges, we consider computational domains belonging  to the class of the elongated domains, boundary conditions at the inner walls (parapets), and inflow conditions that  mimic queue-like entrance processes (cf.~\cite{fujino1993EESD} in  footbridges context).

Therefore, in this section we provide modelling solutions for the following aspects:
\begin{inparaenum}[i)]
\item interactions among individuals;
\item bounded spatial domains $D\subsetneq\R^2$; 
\item pedestrian behaviour at walls;
\item numerical implementation in real geometries;
\item crowd inflow.
\end{inparaenum}

\subsection{Modelling pedestrian interactions}
The interaction velocity $v_i$ (cf.~\eqref{eq:vi}) is the modelling element which accounts for the interaction among pedestrians codified in the kernel $K$. It is worth remarking that, in order to make the model calibration affordable, $K$ should depend on a limited number of  parameters.

We here propose an expression of $K$ based on two parameters, which we calibrate in Section~\ref{sect:application} to produce simulations.

In normal crowd avoidance regime, pedestrian interactions are repulsive and limited to a frontal region (being, thus, \emph{anisotropic}). Therefore, a possible expression of $K$ can be:
\begin{equation}
	K(y-X_t)=-\frac{c}{\max\{\abs{y-X_t},\,R_b\}}\cdot\frac{y-X_t}{\abs{y-X_t}},
	\label{eq:K}
\end{equation}
where $c>0$ is a constant which determines the reference strength of the repulsion and $R_b>0$ is a \emph{body radius}, below which the strength of the repulsion is assumed to be constant and maximum (in particular, $\abs{K(y-X_t)}=c/R_b$ for $\abs{y-X_t}\leq R_b$). We notice that such a cut-off is also necessary in order to avoid any singularity of $K$ for $y=X_t$.

\newcommand{\fontsizehere}{\Large}
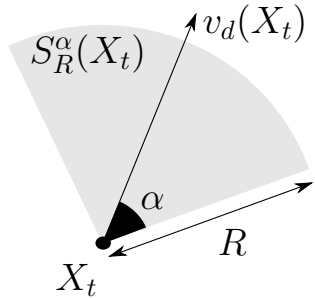
\begin{figure}[t]
\centering
\begin{tikzpicture}
\node[inner sep=0pt] at (-2.3,2){
\includegraphics[width=4cm]{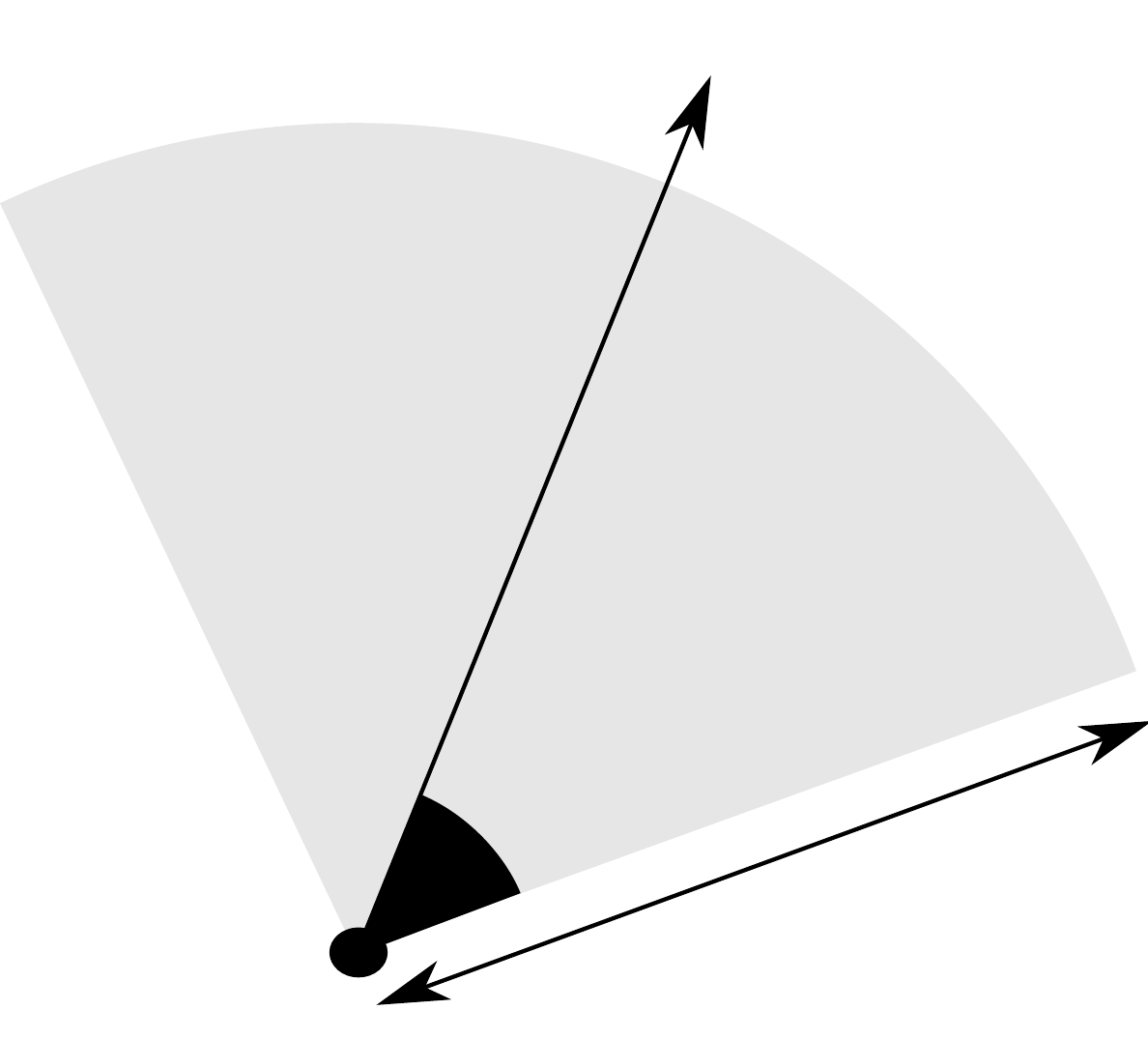}};
\node at (-3.25,3) {\fontsizehere $S^\alpha_R(X_t)$}; 
\node at (-1.05,3.4) {\fontsizehere$v_d(X_t)$};
\node at (-1.35,.5) {\fontsizehere$R$};
\node at (-3.4,0) {\fontsizehere$X_t$};
\node at (-2.4,1.05) {\fontsizehere$\alpha$}; 
\end{tikzpicture}
\caption{Sensory region $S^\alpha_R(X_t)$ of the test pedestrian in $X_t$}
\label{fig:S}
\end{figure}

The interaction described in~\eqref{eq:K} is confined to the sensory region $S(X_t)$, which is modelled as a circular sector of radius $R$, center $X_t$, and angular half-amplitude $0<\alpha<\pi/2$ around $v_d(X_t)$. Formally, the set $S(X_t)$ can be written as
\begin{align*}
	S(X_t)=S^\alpha_R(X_t)&=\Bigl\{y\in D\,:\,\abs{y-X_t}<R, \\
		&v_d(X_t)\cdot\frac{y-X_t}{\abs{y-X_t}}>\vert v_d(X_t)\vert\cos{\alpha}\Bigl\},
\end{align*}
see Fig.~\ref{fig:S}.

It is worth remarking that we choose the orientation of the sensory region according to the desired velocity, which, in principle, can disagree from the  velocity $\dot{X}_t$ adopted by the pedestrian. In this paper we focus on elongated domains, like walkways, on which pedestrians mainly walk from one side to the other. In this setting, the total velocity is not expect to deviate substantially from the desired velocity, thus this modelling approximation appears reasonable. On the other hand, using the actual velocity to define the orientation of the sensory region yields an implicit definition and technical difficulties arise. The interested reader is addressed to~\cite{evers2014anisotropic} where such issues are studied for particle systems having velocity analogous to~\eqref{eq:streamlines}.     

\subsection{Modelling the effects of walls}
\label{sect:walls}
The model we introduced  operates in the  unbounded space $\R^2$. However, domains relevant in applications commonly feature built perimeters, in general \emph{walls}, that agents cannot cross. Therefore, proper behavioural rules describing the agents' reaction to walls should be introduced.

The rules we propose here are obtained on the basis of a pure physical intuition, supported nonetheless by the evidence that both desired and total velocities should not allow the crossing of walls. Constructing suitable desired velocity fields in presence of walls for generic bounded domains is a nontrivial task, thus in the next section we will specifically consider \emph{elongated} domains in view of the application to footbridges.

\subsubsection{Desired velocity}
\label{sect:vd}
In our modelling approach, the  desired velocity field $v_d$, which  drives pedestrian motion in the domain, is supposed to be known \emph{a priori}. Therefore, it should be constructed only from the knowledge of the geometry of the domain.

\begin{figure}[t]
\centering
\includegraphics[width=0.8\textwidth]{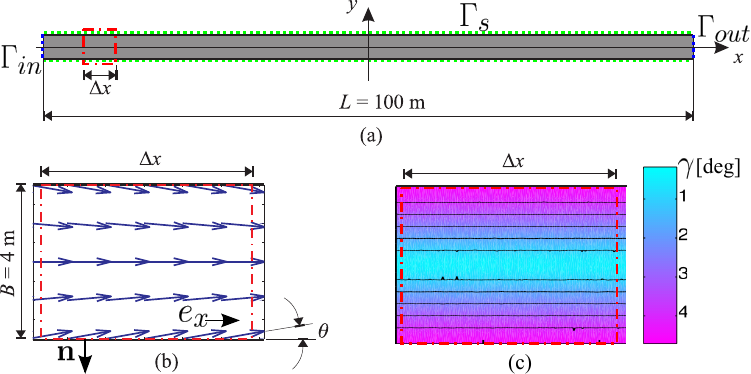}
\caption{(a) Rectangular domain $D$ - (b) Detail of $D$ with the desired velocity vector field. c) Iso-lines of angle $\gamma$}
\label{fig:rectangularD}
\end{figure}

Given a generic wall bounding the domain, having outward normal unit vector $\vect{n}$ (see Fig.~\ref{fig:rectangularD}(b)), $v_d$ must satisfy the following \emph{compatibility} condition:
\begin{equation}
	v_d\cdot\vect{n}\leq 0.
	\label{eq:compat_vd}
\end{equation}

When general domains are considered, constructing a field $v_d$ which is phenomenologically acceptable and satisfies~\eqref{eq:compat_vd} is a nontrivial task, which has been considered both in crowd modelling literature~\cite{piccoli2009CMT,piccoli2011ARMA,maury2010M3AS} and, more generally, when \emph{path planning} is concerned (e.g., robot motion planning~\cite{rimon1992IEEE}). We here propose a method to build this field in case of elongated domains, i.e.~abstractions of footbridges or walkways. These domains are characterized by a longitudinal dimension crossed by pedestrians (\emph{length}) much larger than the transversal dimension (\emph{chord}), which potentially slowly varies.  As in~\cite{maury2010M3AS,piccoli2011ARMA}, we consider velocity fields that are (normalized) potential fields, i.e., having form $v_d=-\nabla{u}/\vert\nabla{u}\vert$ for some potential function $u$  to be determined.

We begin with the simplest domain geometry, where the problem can be easily set and modelling considerations appear more intuitive. Let $D$ be a rectangular domain of length $L$, chord $B$, and aspect ratio $\tilde{B}=B/L=\frac{1}{25}\ll 1$ (Fig.~\ref{fig:rectangularD}(a)). To find the potential $u$, the following assumptions are made:
\begin{itemize}
\item pedestrians flow from left to right in $D$. Therefore, the field $-\nabla u$ must  be directed rightward. Moreover, a \emph{unit} potential difference across $L$ is assumed;
\item pedestrians avoid  ``scratching'' the lateral boundaries when walking. Hence, $-\nabla u$ must be directed inwards at the extrema of the chord and longitudinally at mid-chord. In other words, denoting by $\vect{e}_x$ the unit vector in the longitudinal direction, the angle
\begin{equation*}
	\gamma=\cos^{-1}\left(\frac{v_d\cdot\vect{e}_x}{\vert v_d\vert}\right)
\end{equation*}
is expected to decrease monotonically to zero when approaching  mid-chord. We parametrize the spatial rate at which $v_d$ relaxes toward the horizontal direction in terms of its slope at walls:
\begin{equation}
	\tan{\theta}=\tan{\gamma\vert_{\Gamma_s}},
	\label{eq:tantheta}
\end{equation}
see Fig.~\ref{fig:rectangularD}(b). We stress that, since pedestrian-pedestrian interactions are repulsive, if pedestrian-wall repulsion is not taken into account, agent agglomerations are likely to appear in the proximity of the lateral boundaries $\Gamma_s$.
\end{itemize}

A minimal potential complying with the assumptions above is:
\begin{equation}
	u(x,\,y)=-x+qy^2,
	\label{eq:uparab}
\end{equation}
where the coordinates $x$ and $y$ are scaled with respect to the span $L$ and moreover $q=\tan{\theta}/\tilde{B}$. Hence, $u$ defines the potential field $-\nabla{u}=(1,\,-2qy)$, which, up to normalization, generates the desired velocity field
\begin{equation*}
	v_d=\frac{(1,\,-2qy)}{\sqrt{1+4q^2y^2}}.
\end{equation*}

To deal with more general low-aspect ratio domains, we primarily observe  that~\eqref{eq:uparab} solves the following Poisson problem:
\begin{equation}
	\left\{
	\begin{array}{rcll}
 		\Delta{u} & = & 2q & \textup{in\ } D \\
		\dfrac{\partial u}{\partial\vect{n}} & = & \tan{\theta}\dfrac{\tilde{b}}{\tilde{B}} & \textup{on\ } \Gamma_s \\[3mm]
		u & = & -x_i+qy^2 & \textup{on\ } \Gamma_i\ \textup{($i\in\{in,\,out\}$)},
	\end{array}
	\right.
	\label{eq:poisson}
\end{equation}
which can be imposed on more general domains than the rectangle. As in~\eqref{eq:tantheta}, the term $\theta$ is intended to parametrize the spatial rate at which $v_d$ gets aligned with the longitudinal direction at mid-chord. Therefore, we introduce the factor $\tilde{b}/\tilde{B}$, where  $\tilde{b}=\tilde{b}(x)$ is the (spatially variable) chord amplitude. In rectangular domains, where $\tilde{b}\equiv\tilde{B}$, the Neumann boundary condition on $\Gamma_s$ turns out to be exactly condition~\eqref{eq:tantheta}.

\begin{figure}[t]
\centering
\includegraphics[width=\textwidth]{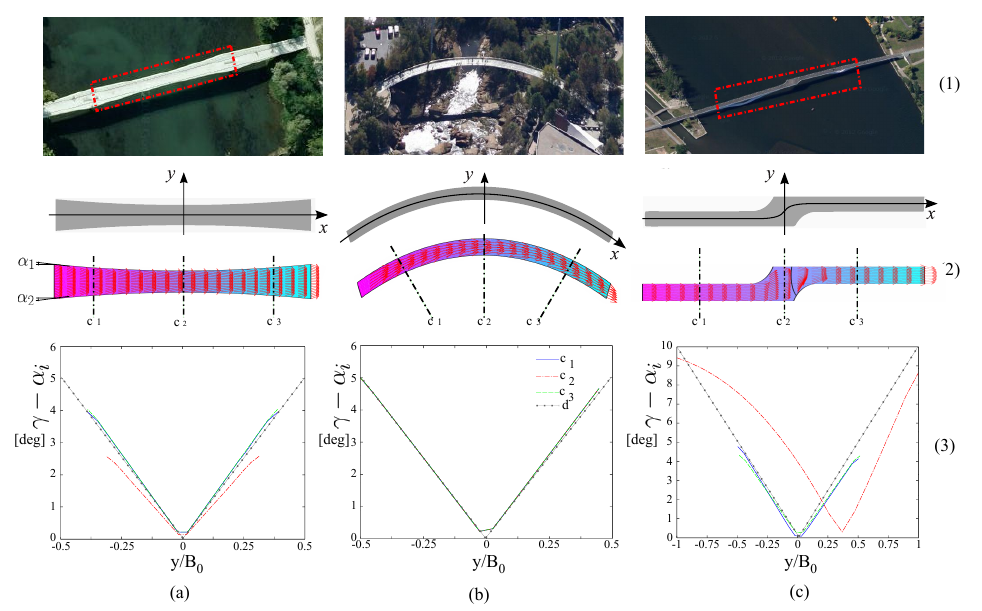}
\caption{(1) Pictures of real footbridges - (2) Computational domains and desired velocity vector fields - (3) Distribution of $\gamma-\alpha_i$ across given chords}
\label{fig:footbridges}
\end{figure}

The potential  in~\eqref{eq:poisson} is subharmonic. In the literature, analogous velocity fields have been usually built by using harmonic functions~\cite{piccoli2011ARMA,rimon1992IEEE,connolly1990IEEE}, relying on the so called \emph{min-max principle}. This principle ensures that any local maximum or minimum of $u$, i.e., a stationary point of the gradient field, lies on the boundary of $D$~\cite{doob2011BOOK}. Therefore, provided proper Dirichlet conditions are imposed (e.g., $u=0$ on exits, $u=1$ anywhere else on the boundary), velocity fields driving pedestrians to a given exit are obtained. However, no control on the behaviour of $-\nabla u$ at walls is allowed, hence phenomenologically unsatisfactory dynamics can arise. Nevertheless, subharmonic functions do not satisfy a minimum principle, hence the absence of internal local minima cannot be guaranteed anymore. Aside from this, their use through~\eqref{eq:poisson} allows us to obtain phenomenologically consistent fields~\cite{iniguez2009PROC}.

In fig.~\ref{fig:footbridges} we applied model~\eqref{eq:poisson} with $\theta=5\degree$  to some real world footbridges having different walkway shapes, respectively: bottleneck shape (Chiaves footbridge~\cite{russell2005BDE}), curved shape (Liberty footbridge~\cite{bogle2004LS}), shifted shape (Coimbra footbridge~\cite{caetano2010ES}). As far as the longitudinal direction is concerned (Fig.~\ref{fig:footbridges}(2)), the obtained $v_d$ fields are constant and a unit potential difference across $L$ is achieved. Concerning the chord-wise direction, for the sake of clarity, the desired walking angle $\gamma$ is referred to the local geometrical inclination of the sidewall. To do so, the angle $\beta:=\gamma-\alpha_i$ is introduced, where $\alpha_i$ extends the geometric angle between the boundary and $\vect{e}_x$. In particular, we express it via the function
\begin{equation*}
	\alpha_i(y)=\left\vert\alpha_1\cdot\frac{y-y_2}{y_1-y_2}-\alpha_2\cdot\frac{y-y_1}{y_2-y_1}\right\vert,
\end{equation*}
where the $\alpha_j$'s ($j=1,\,2$) are defined in Fig.~\ref{fig:footbridges}(2a). In Fig.~\ref{fig:footbridges}(3) the angle $\beta$ is plotted along the sections c1 ($x=-0.35L$), c2 ($x=0$, mid-span), c3 ($x=0.35L$) (Fig.~\ref{fig:footbridges}(2)). It is worth pointing out that the obtained $\beta$ chord-wise profiles are close to the expected trend (gray dotted lines, Fig.~\ref{fig:footbridges}(3)) when the walkway section is almost constant; nonetheless, the $\beta$ profile coherently departs from this trend  when significant geometry variations take place (e.g.~\ref{fig:footbridges}(a)-c2,~\ref{fig:footbridges}(c)-c2).

\subsubsection{Total velocity}
The requirement of geometric compatibility shall reflect also on the total pedestrian velocity at side boundaries. In formulas, in analogy to~\eqref{eq:compat_vd}, it must hold
\begin{equation}\label{eq:gen-wall-impermeability}
	\left.\dot{X}_t\cdot\vect{n}\right\vert_{\textup{walls}}=\left.\left(v_d+v_i\right)\cdot\vect{n} \right\vert_{\textup{walls}}\leq 0.
\end{equation}
Two opposed dynamics complying with~\eqref{eq:gen-wall-impermeability} can be suggested:
\begin{enumerate}
\item an incompatible motion results in a \textit{frictionless} sliding, i.e.,
\begin{equation}
	\left.\dot{X}_t\right\vert_{\textup{walls}}=
		\begin{cases}
			\left(v_d+v_i\right)-\left(\vect{n}\cdot\left(v_d+v_i\right)\vect{n}\right)
				& \text{if\ } \left(v_d+v_i\right)\cdot\vect{n}\geq 0 \\
      		\left(v_d+v_i\right) & \text{otherwise;} 
   		\end{cases}
	\label{eq:scrape}
\end{equation}
\item an incompatible motion results  in a pause, i.e.,
\begin{equation}
	\left.\dot{X}_t\right\vert_{\textup{walls}}=
		\begin{cases}
			0  & \text{if\ } \left(v_d+v_i\right)\cdot\vect{n}\geq 0 \\
      		\left(v_d+v_i\right) & \text{otherwise;}
		\end{cases} 
	\label{eq:impermeability}
\end{equation}
\end{enumerate}
In the simulations presented in Section~\ref{sect:application} we chose the boundary condition~\eqref{eq:scrape}. For an alternative  approach to enforce constraints related to walls based on admissible velocity cones, the interested reader may refer e.g. to~\cite{maury2010M3AS,faure2015crowd}.

\subsection{Simulation of crowd events in articulated domains}
In the current section we deal with  the numerical discretization of~\eqref{eq:strong} for producing numerical simulations of crowd events in real, possibly articulated, domains. We consider spatial discretizations, done via unstructured triangular meshes.

We can approach numerically Problem~\eqref{eq:strong}-\eqref{eq:initcond}  by means of the \emph{ad hoc} scheme proposed in~\cite{piccoli2011ARMA}. We approximate the model equations by means of a twofold discretisation in time and space.
First, a first order explicit-in-time approximation is introduced. Let $[0,\,T]$ be a time interval of interest. We introduce a lattice $t_n= n\Delta t$, with $n=0,\ldots,M$, where $\Delta t$ is such that $\Delta t= T/M$. In this setting, we generate recursively a countable family of densities $\{\tilde{\rho}_n\}_{n=0}^M$, which approximates the $\rho_{t_n}$'s. To this end, we introduce  the discrete-in-time flow map:
\begin{equation}
	\tilde{X}_n(x)=x+\tilde{w}_n(x)\Delta{t},
	\label{eq:Xn}
\end{equation}
where $\tilde{w}_n(x):=v_d(x)+v_i[\tilde{\rho}_n](x)$. After~\eqref{eq:Xn}, we define $\tilde{\rho}_{n+1}$ as the density function  satisfying
\begin{equation}\label{eq-recursive-relation}
	\int_{E}\tilde{\rho}_{n+1}(x)\,dx=\int_{\tilde X_n^{-1}(E)}\tilde{\rho}_n(x)\,dx, \quad\forall E\subseteq D,\  n\geq 0,
\end{equation}
which produces the desired time approximation:
\begin{equation*}
	\tilde{\rho}_n\approx\rho_{n\Delta t}, \quad  n=1,\,2,\,\dots,\,M.
\end{equation*}
If the discrete flow map meets suitable regularity conditions then $\tilde{\rho}_n$ is well-defined as an integrable density for all $n$ (for a detailed discussion on the requirements, the reader can refer to~\cite{piccoli2011ARMA}).

After the time discretization has been established, we consider a finite-volume-type partition of $D$. Let $\{E_k\}_{k=1}^{Q}$ be a grid of $Q$ measurable elements of centroids $x^{k}$. We approximate the density $\tilde{\rho}_n$  by means of a piecewise constant function $\hat{\rho}_n$ which reads
\begin{equation}\label{eq:piecewise-fz}
	\hat{\rho}_n(x)=\sum_{k=1}^{Q}\rho^{k}_n\ind{E_k}(x),
\end{equation}
where $\rho^{k}_n$ is a characteristic value of $\tilde\rho_n$ when restricted to the element $E_k$ (e.g., $\rho^{k}_n=\tilde{\rho}_n(x^{k})$).

Furthermore, we consider a piecewise constant space approximation of the flow map~\eqref{eq:Xn}:
\begin{equation}\label{eq:discrete-flow-map}
	\hat{X}_n(x)=x+\hat{w}_n(x)\Delta{t},
\end{equation}
where we used the piecewise version of the velocity $\hat{w}_n(x)$, i.e.,
\begin{equation*}
	\hat{w}_n(x)=\sum_{k=1}^{Q}(v_d(x^{k})+v_i[\hat{\rho}_n](x^{k}))\ind{E_k}(x).
\end{equation*}
As the mesh is fixed in time, the recursive relation~\eqref{eq-recursive-relation}, once used to define the approximated sequence $\hat\rho_n$ via~\eqref{eq:discrete-flow-map}, can be conveniently tested against the grid elements. From that we produce
\begin{equation*}
	\int_{E_q}\hat{\rho}_{n+1}(x)\,dx=\int_{\hat{X}_n^{-1}(E_q)}\hat{\rho}_n(x)\,dx,
\end{equation*}
which finally yields
\begin{equation}
	\hat{\rho}^{q}_{n+1}=\sum_{k=1}^{Q}\hat{\rho}_n^{k}
		\frac{\vert E_q\cap\hat{X}_n^{-1}(E_k)\vert}{\vert E_q\vert}, \quad q=1,\,2,\,\ldots,\,Q,\  n\geq 0,
	\label{eq:numscheme}
\end{equation}
where $\vert\cdot\vert$ denotes the area of the element. If the spatiotemporal grid is properly refined (viz. under a suitable relationship between the characteristic size of the elements and the time step $\Delta{t}$), the scheme converges to (the weak solution of) Problem~\eqref{eq:strong}-\eqref{eq:initcond} (for technical details see~\cite{piccoli2013AAM,tosin2011NHM}). It is worth to remark that following~\eqref{eq:crowding}, and using~\eqref{eq:piecewise-fz}, we can calculate the (numerically approximated) crowding of a measurable $E\subseteq D$ as
\begin{equation}\label{eq:measure-approx-of-set}
\int_{E}\hat\rho_n(x)\,dx = \sum_{k=1}^{Q}\rho^{k}_n\int_{E}\ind{E_k}(x)\,dx = \sum_{k=1}^{Q}\rho^{k}_n|E \cap E_k|.
\end{equation}

\begin{figure}[t]
\centering
\includegraphics[width=0.8\textwidth, trim = 0cm .15cm 0cm 0cm]{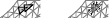}
\caption{Conceptual sketch of the numerical scheme in action}
\label{fig:numscheme}
\end{figure}
   
This numerical scheme can be ideally used with any type of grid, independently of the element shape. From the strict implementation point of view, we need to compute rigid movements of the elements $E_k$ and evaluate their intersections. For instance, in~\cite{cristiani2011MMS,maury2011NHM,piccoli2011ARMA} orthogonal grids formed by square-shaped elements have been used. Nonetheless, when articulated domains are considered, triangular meshes are often used. Pairs of triangles exhibit a wide selection of (topologically) different ways of intersecting\footnote{Notice that intersections of rectangles taken from orthogonal grids, instead, give rise just to rectangular intersections.} (see Fig.~\ref{fig:numscheme} and e.g.,~\cite{sampoli2004TECH}), which makes the evaluation of~\eqref{eq:numscheme} expensive. It is worth pointing out that the problem of intersecting triangles (and convex polygon in general) is typical in computational geometry and computer graphics and  can be solved very efficiently. Algorithms  having linear complexity with respect to the number of edges of the polygons have been conceived (see e.g.,~\cite{orourke1994BOOK,toussaint1983PROC}), which allow computational times to be reduced. We discretised the domains considered in simulations in Section~\ref{sect:application} (see Fig.~\ref{fig:footbridges}) by using triangular meshes.   

\subsection{Pedestrian inflow}
\label{sect:ped_enter}          
In applications, the spatial region of interest is usually just a subset of a larger phenomenological domain in which the crowd moves. Consequently, all people are normally not assumed to be already in the computational domain at the initial time (apart from very specific cases like e.g., evacuation problems~\cite{schadschneider2011EEE}), and the entrance of pedestrians in the domain through an access zone has to be modeled. Pedestrians accessing the domain, even in the simple case of queue-like arrivals, are expected to feature dynamical aspects. Empty and overcrowded access areas provide extreme examples of this aspect: a \textit{free flow}~\cite{venuti2009PLR} is expected in the first case, and (likely) no flux in the second case. Therefore, dynamically adapting boundary conditions appear to be a phenomenological requirement. Dynamic boundary conditions have been explored for instance in the cases of heat-like equations~\cite{vazquez2008heat,vazquez2009laplace}, or of binary fluid separation~\cite{colli2015cahn,miranville2005exponential}. Furthermore, recently an approach to handle flux boundary conditions based on shrinking and absorbing boundary layers has been proposed for one dimensional models of social interactions~\cite{evers2015mild}. However, to the best of our knowledge, just ordinary boundary conditions~\cite{algadhi1990modelling,algadhi1991simulation,colombo2011modelling} or unbounded geometries~\cite{cristiani2011MMS,piccoli2009CMT} have  been considered when deriving mathematical models for crowd movement.

In the current section we propose an approach to allow for pedestrian entrance and arrival, which is conceptually consistent with the model previously developed.
 
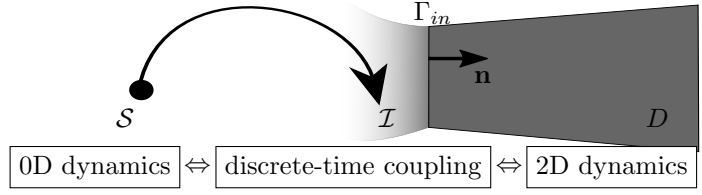
\begin{figure}[t] 
\centering 
\begin{tikzpicture}
\node[inner sep=0pt] (russell) at (0,0)
    {\includegraphics[width=7.5cm,height=2cm]{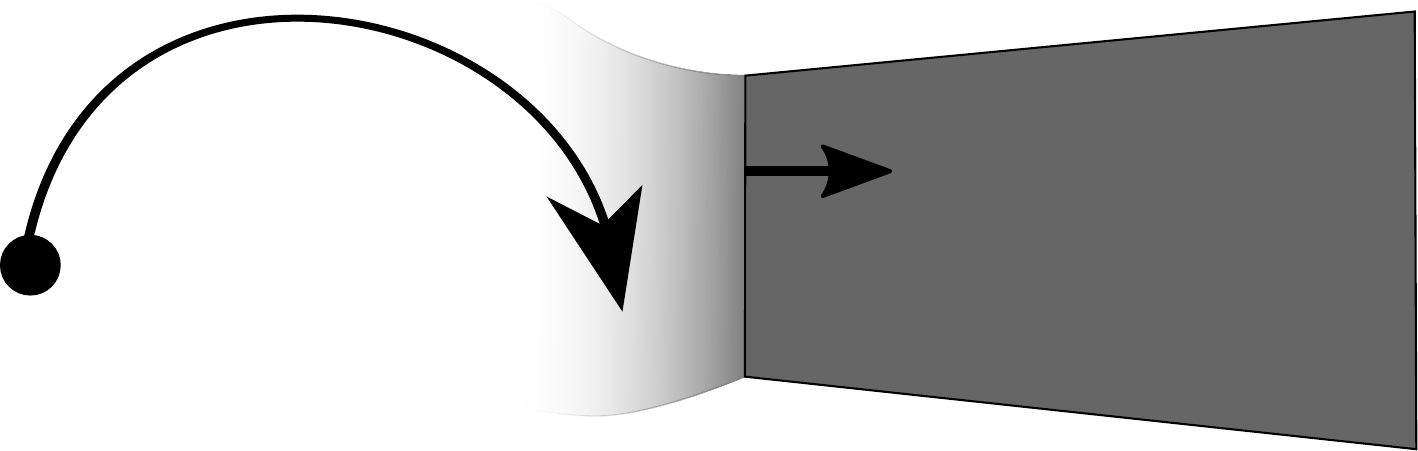}};
\node at (-3.8,-.55) {$\cS$};
\node[draw,fill=white] at (-4.2,-1.2) {0D dynamics};
\node (n1) at (-2.85,-1.2) {$\Leftrightarrow$};
\node at (-.35,-.55) {$\cI$};
\node at (.9,-.0) {$\textbf{n}$};
\node at (.25,.85) {$\Gamma_{in}$};
\node [draw,fill=white] at (-.8,-1.2) {discrete-time coupling};
\node (n2) at (1.25,-1.2) {$\Leftrightarrow$};
\node at (3.2,-.55) {$D$};
\node[draw,fill=white] at (2.6,-1.2) {2D dynamics};
\end{tikzpicture}
\caption{Conceptual scheme of the  entering dynamics} 
\label{fig:SID}
\end{figure}

We consider a group of $N$ pedestrians who want to enter the considered domain. In particular, let $S_t$ be the number of pedestrians still waiting, say in a zero-dimensional ``bulky reservoir'' $\cS$ (cf. Fig.~\ref{fig:SID}), to approach the facility. The quantity $S_t$ evolves in time depending on the number $I_t$ of pedestrians occupying a two-dimensional \textit{entrance} region $\cI$ localized in a boundary layer adjacent to the computational domain $D$. We introduce a function $f:\R\rightarrow\R$ that models the emptying rate of $\cS$, i.e., the entrance rate in $\cI$:
\begin{equation*}
	\dot{S}_t=f(S_t,\,I_t;\,t)
\end{equation*}
along with the initial condition $S_0=N$. We assume:
\begin{itemize}
\item[i.] $f\propto\sigma(S_t)$ for some nonnegative function $\sigma$ satisfying $\sigma(0)=0$, i.e., the entrance rate in $\cI$ is influenced by the amount of people still waiting, and vanishes when all the $N$ pedestrians have flowed into $\cI$.

We further consider an arrival rate in $\cI$ which smoothly decreases when the number of pedestrians in the bulky reservoir $\cS$ is small. Hence we set:
\begin{equation*}    
	\sigma(S_t)= 
	\begin{cases}
		F\frac{S_t}{Np} & \text{if\ } 0\leq S_t/N\leq p \\
		F & \text{if\ } p<S_t/N\leq 1,
	\end{cases}
\end{equation*}
where $F>0$ is a constant and $p\in [0,1]$ is the fraction of $N$ at which the appearance rate starts to decay.
\item[ii.] The entrance region $\cI$ cannot be indefinitely populated. The crowd flow is expected to stop (and possibly revert) after a certain crowding capacity value $C>0$ is reached. Therefore, we  introduce a logistic factor
$$ f\propto\left(1-\frac{I_t}{C}\right). $$ 
The constant $C$  satisfies $C=\rho_C\vert\cI\vert$, $\rho_C>0$ being a given threshold density in $\cI$. Hence, when $I_t<C$ the pedestrian mass flows into $\cI$ and $S_t$ decreases, whereas when $I_t>C$ a reverse flow takes place and $S_t$ increases.
\end{itemize}
Combining the effect of (i) and (ii), we get the following evolution equation for the variable $S_t$
\begin{equation}
	\dot{S}_t=\sigma(S_t)\left(1-\frac{I_t}{C}\right).
	\label{eq:f}
\end{equation}

Obviously, the total number $N$ of pedestrians involved in the crowd event has to be globally conserved in time. In other words, the total mass of people in the bulky reservoir $\cS$ and in the entrance region $\cI$, plus the one flowing into the computational domain, has to be constant in time during the arrival process. Hence we impose the following mass balance:
\begin{equation}
	\dot{S}_t+\dot{I}_t+\Phi_t=0, 
	\label{eq:ISPhi}
\end{equation} 
where 
$$ \Phi_t=\int_{\Gamma_{in}}\rho_t\dot{X}_t\cdot\textbf{n}\,ds $$
is the flux of pedestrian mass transferred across the interface $\Gamma_{in}$ between $D$ and $\cI$ (i.e., $\Gamma_{in}=\partial\cI \cap \partial D$ and $\textbf{n}$ is the unit vector pointing from $\cI$ to $D$). 

\begin{remark}
A similar modelling strategy can be used to describe also the egress of pedestrians from a crowded facility, if one wants to simulate downstream conditions that may affect the outflow of pedestrians (such as e.g., queues). In the present application we refrain from going into such a detail and we assume instead a free outflow of pedestrians from the computational domain.
\end{remark}

In the two-dimensional entrance region $\cI$ pedestrians flowing from $\cS$ are first homogeneously distributed in space, i.e., they are given a constant density across $\cI$, which is then evolved by means of model~\eqref{eq:strong}. In this way it is naturally transported into the computational domain $D$ coherently with the main dynamics taking place there.

The reason why pedestrians from $\cS$ are not directly ``poured'' into $D$ is that in passing from a zero-dimensional to a two-dimensional domain an approximation of their spatial distribution has to be imposed, which, as just stated, here we choose to be the homogeneous one. A ``transition layer'', here represented by the entrance region $\cI$, is then technically necessary to this purpose.

The whole procedure just described is better formalized, in mathematical terms, at a discrete-time level, which is also useful for the numerical implementation of the corresponding equations. More details about this are given in the following.

\subsubsection{Numerical implementation}
\label{sect:entrance_num}
We here extend the numerical scheme~\eqref{eq:numscheme} to treat also the pedestrian arrival process described in Section~\ref{sect:ped_enter}. It is important to stress that model~\eqref{eq:strong} defines a two dimensional dynamics over $D$, conversely~\eqref{eq:ISPhi} prescribes the zero dimensional dynamics of the \textit{number} of entered pedestrians. Therefore,  the region $\cI$ does not only serve as a physical space to give access to the domain $D$; it also allows for the interplay between the two models, which operate in different dimensions.

To couple the models, an explicit-in-time algorithmic procedure based on~\eqref{eq:numscheme} and~\eqref{eq:ISPhi} is here proposed. In particular, in each time step, the crowd distribution in the region $\cI$ is updated in two different stages according alternatively to~\eqref{eq:strong} or to~\eqref{eq:ISPhi}. 

In the current approach, we consider an \emph{extended} spatial domain  $D\cup\cI$, which we discretise via a mesh $\{E_j\}_{j=1}^{Q}$. We require that no mesh element crosses the interface between $D$ and $\cI$, i.e., in formulas,
\[ \abs{E_j \cap D}\cdot\abs{E_j\cap \cI}=0 \qquad \forall j=1,2,\ldots,Q. \]
Consistently with~\eqref{eq:piecewise-fz}, we consider the discrete density $\hat{\rho}_n^j$ on the  extended domain. Furthermore, for any discrete instant of time $n$, we define
\begin{equation}\label{eq:hatIn}
	I^{\hat{\rho}}_{n}:=\sum_{j\,:\,E_j\subset\cI}\hat\rho^{j}_{n}\vert E_j\vert,
\end{equation}
that measures the total mass contained in the region $\cI$. On side of this discrete-in-time macroscopic variable, we introduce the quantities $S_{n}$ and $I_n$, which approximate, at discrete times, the state variables in~\eqref{eq:f}. Algorithm~\ref{alg:entrance} describes the proposed dynamics of the crowd through $\cS$, $\cI$, and $D$.

\begin{algorithm}[H]
\caption{Complete simulation loop inclusive of entering process\vspace{6pt}}
\SetKw{Evo}{Evolve}
\SetKw{Eval}{Evaluate}
\SetKw{Ass}{Assign}
\SetKw{Com}{Compute}
\SetKwFor{Stage}{}{}{}
\SetAlgoSkip{\bigskip}
\BlankLine
\For{$n\leftarrow 1$ \KwTo $M$}{
\Stage{\textit{Evaluate the mass distribution over the extended domain $D \cup \cI$}}{
\Com $\{\hat{\rho}^{j}_{n}\}_{j=1}^Q$ over $D \cup \cI$ from $\{\hat{\rho}^{j}_{n-1}\}_{j=1}^Q$ via~\eqref{eq:numscheme}
\Com $I^{\hat{\rho}}_n$ from $\{\hat{\rho}^{j}_{n}\}_{j=1}^Q$ via~\eqref{eq:hatIn}
}
\Stage{\textit{Evaluate the new values $S_{n}$ and $I_{n}$}}{
\Com the flux from $\cS$ to $\cI$ (LHS of~\eqref{eq:f})
\vspace{-6pt}
\begin{equation*}
	f_{n} = f(S_{n-1},\,I^{\hat{\rho}}_{n};\,t_n)
	\vspace{-6pt}
\end{equation*} 
\Com $S_{n}$ via $S_{n}=S_{n-1}+\Delta{t}f_n$ \\
\Com $I_{n}$ via $I_{n}=I^{\hat{\rho}}_{n}-\Delta{t}f_n$
}
\Stage{\textit{Update the mass distribution on $\cI$ to conform with $I_{n}$}}{
\Ass
\begin{equation}\label{eq:uniformly-assign}
	\hat{\rho}_{n}^{j}\gets\frac{I_{n}}{\vert\cI\vert}, \quad \forall\ j\,:\,E_j\subset\cI
\end{equation}
}
}
\label{alg:entrance}
\end{algorithm}
It is worth stressing that in the last assignation step~\eqref{eq:uniformly-assign} of Algorithm~\ref{alg:entrance} a transition from the zero-dimensional value $I_n$ of model~\eqref{eq:ISPhi} to a two-dimensional distribution is performed. To cope with the missing information about the spatial distribution of the mass, homogeneous density is enforced.

\section{Application}
\label{sect:application}
In this section we provide some real world applications of the proposed model. In particular, model sensitivity and calibration are discussed in the reference configuration in Fig.~\ref{fig:rectangularD}. Moreover, we discuss simulation results for the geometrical configurations in Fig.~\ref{fig:footbridges}.

\subsection{Setup overview and some simulated phenomena}
We here consider a prototypical crowd event happening in the straight rectangular walkway $D$ depicted in Fig.~\ref{fig:rectangularD}. In this case study we assume that a hypothetical client is able to provide some data of the problem setup, particularly $L$, $B$, the expected total number $N$ of incoming pedestrians, capacity density $\rho_C$ in~\eqref{eq:f}, the pedestrian desired speed $V=\vert v_d\vert$. Here, we selected values on the basis of data available in Transportation and Civil Engineering literature~\cite{buchmueller2006ETH,fujino1993EESD,venuti2007CRM}, see Tab.~\ref{tab:param}. The length $L$ and the desired speed $V$ are reference quantities, thus the time scale $T=L/V$ can be defined as the time required to an undisturbed pedestrian to cross the whole facility.  

\begin{table}
\caption{Parameters used for crowd event simulations} 
\label{tab:param}
\begin{center}
\begin{tabular}{|ccccc|}
\hline
$L$ & $B$ & $\rho_C$ & $N$  & $V$ \\
$100~\unit{m}$ & $4~\unit{m}$ & $1.3~\unit{ped/m^2}$ & $1500~\unit{ped}$ & $1.18~\unit{m/s}$ \\
\hline 
$c^\star$ & $R$ & $\alpha$ & $R_b$ & $\theta$ \\
$5\expten{-4}$ & $2~\unit{m}$ & $45\degree$ & $0.3~\unit{m}$ & $5\degree$\\
\hline
\end{tabular}
\end{center}
\end{table}

\begin{figure}[t]
\centering
\includegraphics[width=\textwidth]{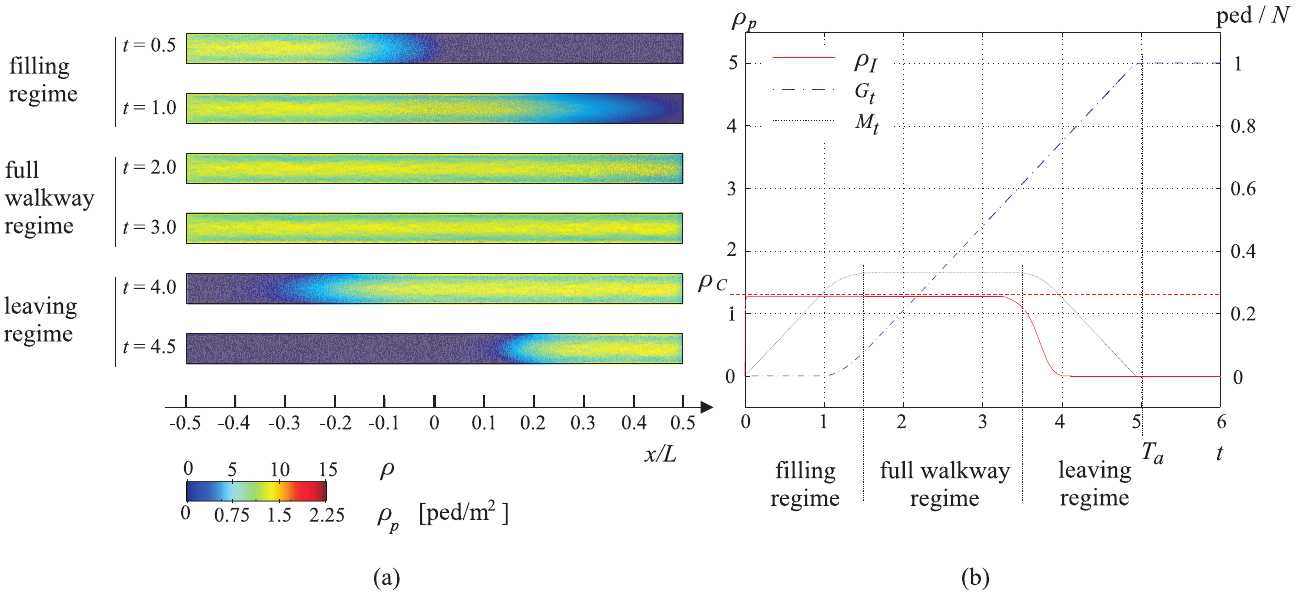}
\caption{(a) Instantaneous density fields and recognized regimes - (b) Time history of some crowd bulk parameters}
\label{fig:regimes_bulk}
\end{figure}

To outline the main features of the simulated crowd event, Fig.~\ref{fig:regimes_bulk}(a) reports some instantaneous density fields and groups them in three recognized regimes: during the \emph{filling regime} pedestrians advance on the partially empty walkway, which is homogeneously filled in the \emph{full walkway regime}. The crowd event gradually ends during the \emph{leaving regime}. In Fig.~\ref{fig:regimes_bulk}(b), the time history of two bulk parameters of the event is also plotted: the number $M_t$ of pedestrians along the walkway and the cumulative number $G_t$ of pedestrians that reached the end and left, both scaled with respect to $N$. A further bulk parameter can be easily recognized, i.e. the \emph{total time} of the crowd event $T_a$ defined as
\begin{equation}
	T_a=\inf\left\{t\colon\frac{G_t}{N}=1\right\}.
	\label{eq:Ta}
\end{equation}

\subsubsection{Sensitivity to free model parameters}
Five free model parameters remain and determine the evolution of the solution. They are the constants $c$, $\theta$, $R$, $R_b$, and $\alpha$. On the one hand, we fix $R=2~\unit{m}$, $R_b=0.3~\unit{m}$, and $\alpha=45\degree$ which, depending on the geometry of the sensory region, we consider case-independent and recoverable from existing literature (see e.g.,~\cite{fruin1987BOOK,venuti2007CRM}). On the other hand, we inquire the sensitivity of the model to the repulsion constant $c$ and to the angle $\theta$ . In particular we focus on the dimensionless version of $c$, namely $c^\star= c/(VL)$, to perform a unit-free calculation, valid for different values of $V$ and $L$ The analysis is based on the  variables $T_a$, which has been previously defined in~\eqref{eq:Ta}, and on $\delta\rho$, i.e.
\begin{equation}\label{eq:delta-rho}
	\delta\rho=\frac{\rho_m-\rho_s}{\rho_C},
\end{equation}
where $\rho_s$ and $\rho_m$ are respectively the crowd density at the walkway side and at the mid-line evaluated at  mid-span ($x=L/2$) during the full walkway regime. In other words, the variable~\eqref{eq:delta-rho}  measures  the chord-wise uniformity of the crowd density, being the span-wise uniformity assured during the full walkway regime and the chord-wise symmetry of the solution assured by the selected setup.

\begin{figure}[t]
\centering
\includegraphics[width=0.8\textwidth]{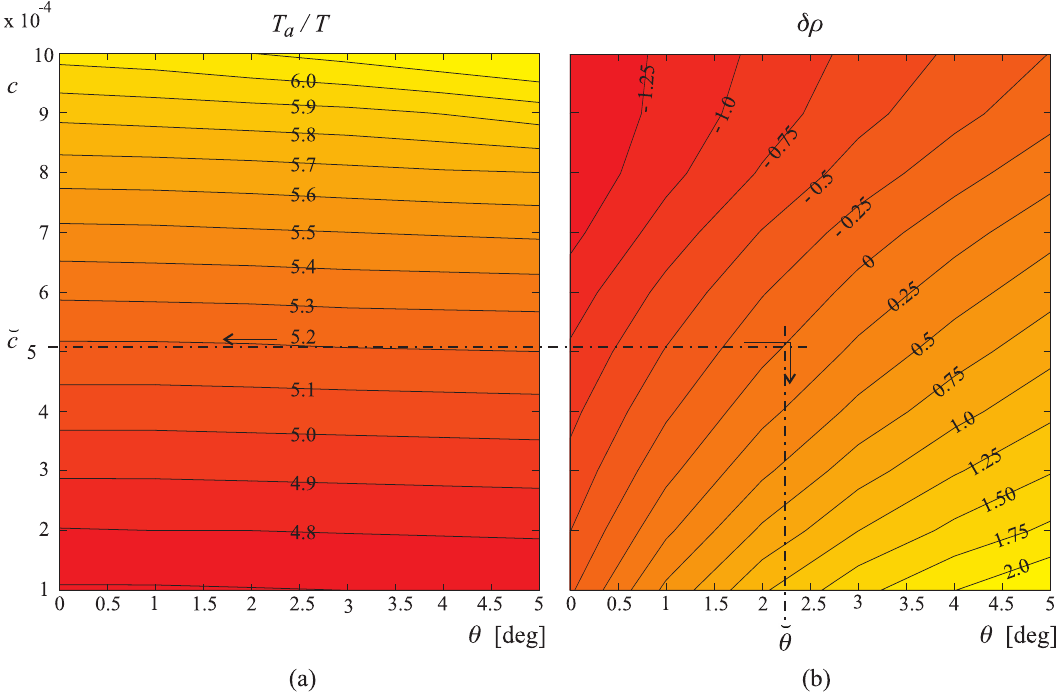}
\caption{(a) Contours of $T_a=T_a(c,\,\theta)$ - (b) $\delta\rho=\delta\rho(c,\,\theta)$}
\label{fig:sensitivity}
\end{figure}

In Fig.~\ref{fig:sensitivity}, we plot these variables  versus the dimensionless parameters $2.5\expten{-4}\leq c^\star\leq 12.5\expten{-4}$, $0\degree\leq\theta\leq 5\degree$.

\begin{figure}[t]
\centering
\includegraphics[width=\textwidth]{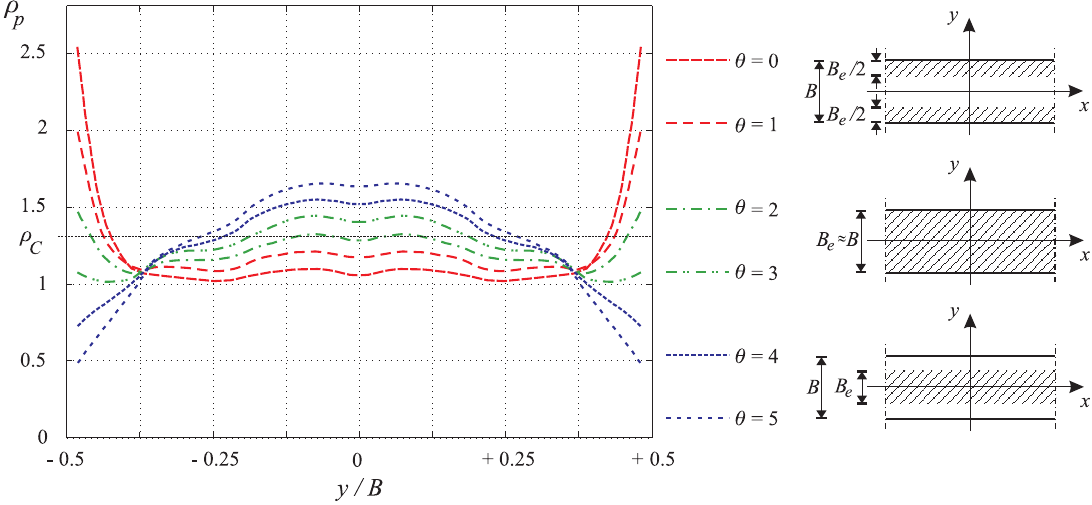}
\caption{Chord-wise crowd density profiles at mid-length for different $\theta$ ($c^\star=5\expten{-4}$)}
\label{fig:chordwise}
\end{figure}

The crowd event time $T_a$ (Fig.~\ref{fig:sensitivity}(a)) is mainly sensitive to the pedestrian-pedestrian repulsion, i.e., to $c^\star$. For the selected incoming pedestrian density $\rho_C$, $T_a$ approximately varies from three to four times the undisturbed pedestrian crossing time, in dependence of $c^\star$. On the contrary, $\delta\rho$ (Fig.~\ref{fig:sensitivity}(b)) depends on both parameters and shows both positive values (higher density at the walkway sides) and negative ones (high density along the mid-line) in the selected range of the model parameters. In other words, for any considered value of $T_a$, there exist a value $\theta^\ast$ of the angle allowing a nearly homogeneous chord-wise crowd density ($\delta\rho=0$), while larger or smaller values produce nonuniform distributions. To detail the chord-wise trend of the crowd density and to discuss its phenomenological features, in Fig.~\ref{fig:chordwise} we report the $\rho$ profiles at mid-length for different values of $\theta$ and  fixed value of $c^\star=5\expten{-4}$.

For $0\leq\theta<\theta^\ast$, where $\theta^\ast\approx 2\degree$ with $c^\star=5\expten{-4}$ we have that 
\begin{inparaenum}[i)]
\item the pedestrian-wall repulsion dictated by the desired velocity field is lower than the pedestrian-pedestrian repulsion;
\item the chord-wise component of the total velocity field is directed toward the lateral walls;
\item the crowd density results larger at the walkway sides than at mid-chord (Fig.~\ref{fig:chordwise}, red profiles).
\end{inparaenum} 
Conversely, the pedestrian-wall repulsion predominates over the pedestrian-pedestrian repulsion for $\theta^\ast<\theta\leq 5\degree$, and the crowd density at mid-chord is larger than at walkway sides (Fig.~\ref{fig:chordwise}, blue profiles). The crowd density profile is almost flat around the value $\theta^\ast$ (Fig.~\ref{fig:chordwise}, green profiles). 

In a general modelling perspective we notice that
\begin{inparaenum}[i)]
\item the parameter $\theta$ allows one to account for different degrees of repulsion of either both or a single walkway side, e.g., a panoramic point on one side only;
\item the parameter $c^\star$ allows one to account for different attitudes of pedestrians in accepting the proximity of other walkers, e.g., dictated by different travel purposes~\cite{buchmueller2006ETH,venuti2007CRM}.
\end{inparaenum}
From a Transportation Engineering perspective, the parameter $\theta$ allows one to model the \textit{shy distance} of pedestrians from the wall~\cite{pushkarev1975BOOK} and to evaluate the \emph{effective width} $B_e$ of the walkway (e.g.,~\cite{habicht1984JTE}): for $\theta=\theta^\ast$ the effective width of the walkway matches the geometric width, while shorter effective widths are obtained otherwise (Fig.~\ref{fig:chordwise}, conceptual sketches).

\subsubsection{Calibration of free parameters}\label{sect-calibration-of-free-param}
The two free parameters considered, i.e. $\theta$ and $c^\star$, characterize the desired and interaction velocity, thus account for distinct avoidance phenomena, respectively pedestrian-wall and pedestrian-pedestrian avoidance. To calibrate these parameters, bulk experimental data directly obtained from real world crowd events are preferable to measurements at the pedestrian scale obtained in laboratory tests. In fact, bulk data allow a calibration procedure that accounts environmental effects (e.g.  travel purpose of pedestrians) and that can be easily applied in the engineering practice. We assume that a hypothetical client (e.g. the footbridge owner or operator) is able to complement the setup data with bulk measurements obtained on the walkway of interest or on analogous facilities. In particular, the crowd event time $T_a$ is easy to measure, hence it is usually provided. For instance, in the following, we adopt $\breve{T_a}/T=5.2$  accordingly to the  measurements \textit{in situ} in~\cite{fujino1993EESD}. Moreover, we require information about the degree of repulsion of lateral sides. It can  either be qualitatively identified by referring to the prototypical blue-red-green profile shapes in Fig.~\ref{fig:chordwise} or it can be quantitatively expressed by $\delta\rho$. For instance, in the following $\breve{\delta}\rho=0$ (flat density profile as in~\cite{fujino1993EESD}) is adopted.

On the basis of the bulk data above and the previous sensitivity analysis, we here suggest a calibration strategy for the model parameters. In engineering terms, the graphs of Fig.~\ref{fig:sensitivity} can be used as ``calibration charts'': first, we obtain the value of the constant $c^\star=\breve{c}^\star$  by setting the value of $T_a/T=\breve{T_a}/T$ (leftward arrow in Fig.~\ref{fig:sensitivity}-a); once $\breve{c}^\star$ is known, we recover the value of $\theta=\breve{\theta}$  by setting the value of $\breve{\delta} \rho$ (downward arrow in Fig.~\ref{fig:sensitivity}-b).

In the following section we retain the values $\breve{c}^\star\approx5\expten{-4}$ and $\breve{\theta}=\theta^\ast\approx 2\degree$, obtained by means of the outlined procedure from $\breve{T_a}/T=5.2$ and $\breve{\delta}\rho=0$. 

\subsection{Crowd flow along real world footbridges}
We here present the result of the simulations of the crowd flow along the same real world footbridges described in Section~\ref{sect:vd}, Fig.~\ref{fig:footbridges},  in order to show the ability of the proposed approach to face actual engineering problems. The simulations adopt the same setup (Tab.~\ref{tab:param}) and the values of the model parameters set in Section~\ref{sect-calibration-of-free-param}, so that the comparison with the reference test-case (rectangular walkway) allows one to qualitatively point out the effects of the domain geometry.

\begin{figure}[t]
\centering
\includegraphics[width=\textwidth]{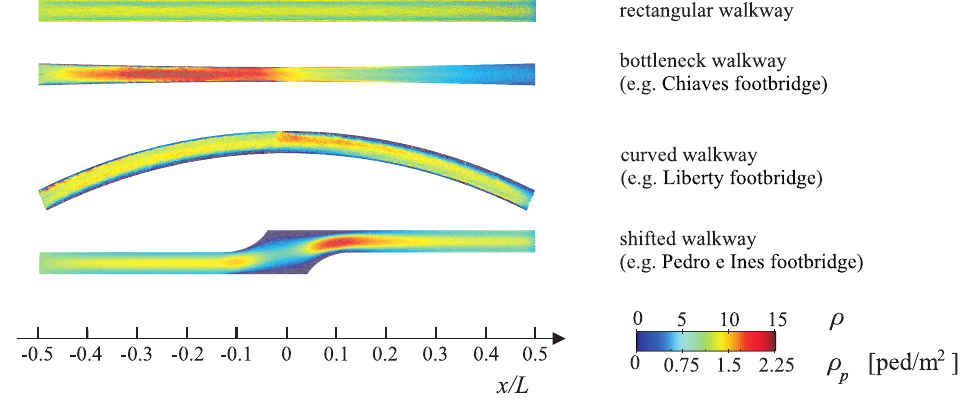}
\caption{Comparison among real world footbridges: density field during the full walkway regime}
\label{fig:comp_footbridges}
\end{figure}

For instance, Fig.~\ref{fig:comp_footbridges}(a) collects the crowd density instantaneous fields during the full walkway regime for the considered domains.

The fields differ significantly both qualitatively and quantitatively: asymmetric patterns arise with respect to the chord-wise axis at midspan (bottleneck walkway), to the longitudinal axis (curved walkway) or to both (shifted walkway); the maximum crowd density nearly doubles in the bottleneck and shifted walkway with respect to the rectangular one.

\section{Discussion}
\label{sect:conclusions}
In this paper, we considered the macroscopic model introduced in~\cite{cristiani2011MMS,piccoli2009CMT,piccoli2011ARMA} to evaluate the collective evolution of crowds on footbridges. We reviewed its derivation from a phenomenological perspective, centred around the motion of the single individual.  Then, we provided a mathematical procedure  to obtain, out of the equations governing the motion of single individuals, an equation describing the collective evolution of the crowd. In view of the issues raised in Section~\ref{Sec:footbr-intro}, we addressed the problems of the discretisation of the equations and of the conditions to impose on lateral walls and on the open boundaries.  Specifically, we proposed a model based on a Poisson problem aimed at deducing phenomenologically-consistent desired velocity fields in generic elongated geometries. This method proved to be successful in the case studies considered yielding velocity fields with the proper direction and expected inward orientation, also in case of curved or out-of-axis facilities.  However, it remains open the problem of finding proper geometric conditions for the   domain such that the sub-harmonic solution to the Poisson problem provides the desired results (e.g.~no local maxima). We further proposed to model the arrival of pedestrians via a dynamic system, that enables a dependence of the pedestrian inflow on the local crowding of the entrance region, as well as on the amount of pedestrians still waiting to appear in the facility. We suggested an algorithmic procedure to make use of this dynamical system at discrete time. Finding a proper analytic and modelling way to formulate the coupling between the dynamical system and the crowding model at continuous time is an open issue as well. 

Considering a simple elongated rectangular domain, we finally performed a sensitivity analysis based on bulk descriptors, namely the total time of a crowd event and the degree of homogeneity of the chord-wise crowd profile. This way, we inquired the impact of selected parameters owing to the microscopic scale on the macroscopic flow. On the basis of the model sensitivity to the parameters, we devised a tuning procedure based on macroscopic measurement \textit{in situ}, which can be possibly read as a  simple ``calibration chart''. To show the feasibility of the proposed approach in actual engineering problems, we simulated crowd events in different computational domains inspired by real footbridges.

\bibliographystyle{plain}
\bibliography{BlCaTa-crowd_footbridges}
\end{document}